\documentclass[11pt]{article}

\usepackage{latexsym,amssymb,amsfonts,graphicx,makeidx,a4wide,theorem,rotating,
ifthen,amsmath,subfigure,epsfig,nicefrac,enumerate}

\newtheorem{theorem}{Theorem}[section]
\newtheorem{example}[theorem]{Example}
\newtheorem{corollary}[theorem]{Corollary}

\newtheorem{definition}[theorem]{Definition}
\newtheorem{problem}[theorem]{Problem}
\newenvironment{proof}{\noindent{\bf Proof~}}{\null\hfill $\Box$\par\medskip}

\newcommand{\graph} {\text{val}}
\newcommand{\val} {\text{val}}
\newcommand{\CW} {\text{CW}}

\newcommand{\NLC} {\text{NLC}}

\newcommand{\MSOA}{\text{MSO}_1}

\newcommand{\cw} {\text{clique-width}}
\newcommand{\nlcw} {\text{NLC-width}}
\newcommand{\lab} {\text{lab}}

\newcommand{\cws} {\text{cw}}
\newcommand{\nlcws} {\text{nlcw}}
\newcommand{\rws} {\text{rw}}
\newcommand{\tws} {\text{tw}}

\begin{document}

\title{Graph Operations on Clique-Width Bounded Graphs}

\author{
Frank Gurski \\
University of D\"usseldorf, Institute of Computer Science\\
Algorithmics for Hard Problems Group, D-40225 D\"usseldorf\\
\texttt{frank.gurski@hhu.de}
}


\maketitle



\begin{abstract}
Clique-width is a well-known graph parameter. 
Many NP-hard graph problems admit polynomial-time solutions when restricted to 
graphs of bounded clique-width. The same holds for NLC-width.
In this paper we study the behavior of clique-width and NLC-width
under various graph operations and graph transformations. 
We give upper and lower bounds for the clique-width and NLC-width 
of the modified graphs in terms of the clique-width and NLC-width 
of the involved graphs.

\bigskip
\noindent
{\bf Keywords:} clique-width, NLC-width,  graph operations,  graph transformations
\end{abstract}


\section{Introduction}\label{sec-intro}

A {\em graph parameter} is a function that associates with every graph a positive integer.
One of the most famous graph parameters is tree-width, which was defined 
by Robertson and Seymour in \cite{RS86}. See \cite{Bod98} for an 
overview on tree-width. Tree-width bounded graphs are  interesting from an 
algorithmic point of view since several NP-complete graph problems can be solved in 
polynomial time on graph classes of bounded tree-width using dynamic programming 
\cite{Arn85,AP89,Hag00,KZN00}.

A further well known graph parameter is clique-width which was defined by 
Courcelle and Olariu in \cite{CO00} through a composition mechanism for
vertex-labeled graphs. 
The NLC-width of a graph was defined by Wanke in
\cite{Wan94} by a composition mechanism similar to that for clique-width.
Both parameters are more powerful than tree-width, since the clique-width and
NLC-width of a graph can be bounded in its tree-width, but not vice versa. 

Clique-width and NLC-width bounded graphs are also interesting from an 
algorithmic point of view. Several NP-complete graph problems can be solved in 
polynomial time on graph classes of bounded clique-width.
For example, all graph properties which are expressible in monadic second order 
logic with quantifications over vertices and vertex sets ($\MSOA$-logic) are 
decidable in linear time on clique-width bounded graphs 
which are  given with an appropriate clique-width $k$-expression \cite{CE12,CMR00}.
Also by using fly-automata problems expressible in $\MSOA$-logic
can be solved if the graphs are given with a $k$-expression \cite{CD12}.
Furthermore, there are also a lot of NP-complete graph problems which are not 
expressible in $\MSOA$-logic like Hamiltonicity, partition problems, and
bounded degree subgraph problems but which can also be solved in polynomial 
time on clique-width bounded graphs \cite{CT10,EGW01a,GW06,KR03,ST07,Wan94}.
In order to apply these algorithms non-optimal expressions are sufficient.
Such expressions can be found by the result shown in \cite{OS06}: For 
every fixed $k$ for every given graph $G$ one
can compute in polynomial time a clique-width $g(k)$-expression or assert that
the clique-width of $G$ is greater than $k$. 

Distance-hereditary graphs have clique-width at most $3$ \cite{GR00}. 
The set of all graphs of clique-width at most $2$ or NLC-width $1$ is the 
set of all co-graphs, i.e. $P_4$-free graphs. Brandst\"adt et al. have analyzed the 
clique-width of graphs defined by forbidden induced one-vertex extensions of 
$P_4$ \cite{BDLM05}. The clique-width and NLC-width of permutation graphs, 
interval graphs, grids, and planar graphs is not bounded \cite{GR00}. 
Every graph of tree-width at most  $k$ has clique-width at most 
$3\cdot 2^{k-1}$ \cite{CR05}. See \cite{KLM09} for a survey on the clique-width 
of graph classes.

The recognition problem for graphs of clique-width
or NLC-width at most $k$ is still open for $k\ge 4$ and $k \geq 3$, respectively. 
The problem whether a graph has clique-width at most $3$ is decidable in 
polynomial time \cite{CHLRR12} and the problem whether a graph has 
NLC-width at most $2$ is also decidable in polynomial time \cite{Joh00,LMR07}. 
By the characterization in terms of co-graphs, it can be decided in
linear time whether a graph has clique-width at most $2$ or NLC-width $1$ 
\cite{CPS85}. 
Computing NLC-width and computing clique-width is 
NP-hard \cite{GW07b,FRRS09}. But the clique-width of tree-width bounded 
graphs is computable in linear time \cite{EGW03}.
An approach to determine the clique-width using an encoding 
to propositional satisfiability (SAT) which is evaluated by a SAT solver
was presented in \cite{HS15}. This approach was extended by a 
combinatorial characterization of clique-width in \cite{CHMPR15}.

A {\em graph transformation} $f$ is a transformation that creates a new graph $f(G_1,\ldots,G_n)$ 
from a number of $n\geq 1$ input graphs $G_1,\ldots G_n$. Examples are taking an induced 
subgraph of a graph, adding an edge to a graph, and generating the join of two graphs.
A {\em graph operation} is a graph transformation which is deterministic and invariant
under isomorphism. Examples are the edge complementation of a graph 
and generating the join of two graphs.\footnote{Please note that by our definition the two graph
transformations taking an induced 
subgraph of a graph and adding an edge to a graph are no graph operations.}
The graph theory books by Bondy and Murty \cite{BM76} and by Harary \cite{Har69} 
include a large number of transformations on graphs.


The impact of graph operations which can be defined by monadic second order formulas 
(so-called MS transductions) on graph parameters can often be shown in a very short way 
although the bounds are rough ones \cite{Cou06,CE12}. 

Transformations that reduce graphs can be used to characterize sets of graphs by forbidden 
graphs.
The property that a graph has tree-width at most $k$ is preserved under the transformation 
taking minors, which is used to show that the set of graphs of tree-width at most $k$ can 
be characterized by a finite set of forbidden minors \cite{RS85}. 

Oum and Seymour introduced in \cite{OS06} the rank-width of graphs, which is defined 
independently of vertex labels, but which is shown to be as powerful as clique-width.
In \cite{Oum05a} it is shown that the property that a graph has rank-width at most $k$ is  
preserved under the transformation taking local complementation, which leads to a 
characterization of graphs of rank-width at most $k$ by finitely many
forbidden vertex-minors (i.e. taking induced subgraphs and 
local complementations). 

It is still open if there exists a graph transformation that does not increase NLC-width or 
clique-width and which can be used to characterize graphs of NLC-width at most $k$ or 
clique-width at most $k$ by a set of finitely many forbidden subgraphs. Such characterizations
would lead polynomial time  recognition algorithms for the corresponding graph classes.


%
%

The effect of graph transformations on  graph 
parameters is well studied, e.g. for band-width in \cite{CO86}, for tree-width in \cite{Bod98}, 
for clique-width briefly in \cite{Cou14,HOSG08}, 
and for rank-width in \cite{HOSG08}.
The behavior of clique-width and NLC-width under various graph operations 
is considered in this paper, which is organized as follows.
In Section \ref{intro}, we recall the definitions of clique-width and NLC-width.
In Section \ref{sec-bin}, we give an overview on the effect of the binary transformations 
join, 
disjoint union, 
union, 
products, 
corona, 
substitution, and
1-sum on the clique-width and NLC-width of given graphs. 
In Section \ref{sec-un}, we consider the latter problem for
the unary graph transformations
quotient,
subgraph, 
edge complement, 
bipartite edge complement, 
power of graphs,
line graphs,
local complementation, 
switching, 
Seidel complementation
edge  addition, 
edge subdivision, 
vertex identification, and
vertex addition.
For the transformations local complementation and
Seidel complementation we even can bound
the clique-width and NLC-width of every graph which is equivalent 
to a given graph, i.e. every graph which can be obtained
by applying an arbitrary number of one of these transformations. 
In Section \ref{sec-con}, we summarize our results,
give extensions to directed and linear versions of clique-width and NLC-width,
some conclusions, and an outlook.

\section{Preliminaries}\label{intro}

\paragraph{Graphs}
We work with finite undirected {\em graphs} $G=(V_G,E_G)$,
where $V_G$ is a finite set of {\em vertices} and $E_G \subseteq \{ \{u,v\} \mid u,v \in
V_G,~u \not= v\}$ is a finite set of {\em edges}.\footnote{Thus we do not consider 
graphs with loops or multiple edges.}
For a vertex $v\in V_G$ we denote by $N_G(v)$ 
the set of all vertices which are adjacent to $v$ in $G$, i.e. 
$N_G(v)=\{w\in V_G~|~\{v,w\}\in E_G\}$. 
Vertex set $N_G(v)$ is called the set of all {\em neighbors} of $v$ in $G$ 
or {\em neighborhood} of $v$ in $G$. Please note that $v$ does not belong 
to $N_G(v)$.
The {\em degree} of a vertex $v\in V_G$,  denoted by $\deg_G(x)$, is 
the number of neighbors of vertex $v$ in $G$, i.e. $\deg_G(v)=|N_G(v)|$.
We are discussing graphs only up to isomorphism. This allows us to
define the path on $n$ vertices 
$P_n=(\{v_1,\ldots,v_n\},\{\{v_1,v_2\},\ldots, \{v_{n-1},v_n\}\})$, which
will be useful in several examples. For the definition of further
special graphs we refer to the book of Brandst\"adt et al. \cite{BLS99}.

\paragraph{Labeled Graphs}
In order to define clique-width and NLC-width, 
we need finite undirected labeled {\em graphs} $G=(V_G,E_G,\lab_G)$,
where $V_G$ is a finite set of {\em vertices} labeled by some mapping
$\lab_G: V_G \to [k]$ and $E_G \subseteq \{ \{u,v\} \mid u,v \in
V_G,~u \not= v\}$ is a finite set of {\em edges}.
The labeled graph consisting of a single vertex labeled by $a \in
[k]$ is denoted by $\bullet_a$. Most of the definitions for unlabeled
graphs can be applied to labeled graphs. Thus, we just want to mention
subgraphs and isomorphism for labeled graphs.

A labeled graph  $J=(V_J,E_J,\lab_J)$ is a {\em subgraph} of $G$ if $V_J
\subseteq V_G$, $E_J \subseteq E_G$ and $\lab_J(u)=\lab_G(u)$ for all $u \in
V_J$. $J$ is an {\em induced subgraph} of $G$ if additionally $E_J=\{ \{u,v\} \in E_G
\mid u,v \in V_J\}$. 
Two labeled graphs $G$ and $J$ are {\em isomorphic} if there is a bijection 
$f:V_G\to V_J$ that preserves adjacencies and the labelings, i.e. 
$\{u,v\} \in E_G \Leftrightarrow \{f(u),f(v)\} \in E_J$ and
$\lab_{G}(u)=\lab_J(f(u))$ for all $u\in V_G$.

\paragraph{Clique-width}
The notion of clique-width\footnote{The operations in the definition of
clique-width were first considered by Courcelle, Engelfriet, and Rozenberg 
in \cite{CER93}.} for labeled  graphs is defined by Courcelle 
and Olariu in \cite{CO00} as follows.
\begin{definition}[Clique-width, \cite{CO00}]
Let $k$ be some positive integer. The class $\CW_k$ of labeled graphs is
recursively defined as follows.
\begin{enumerate}
\item
The single vertex graph $\bullet_a$ for some $a \in [k]$ is in $\CW_k$.
\item
Let $G=(V_G,E_G,\lab_G) \in \CW_k$ and $J=(V_J,E_J,\lab_J) \in \CW_k$ be two
vertex-disjoint  labeled graphs, then $$G \oplus J:=(V',E',\lab')$$
defined by $V':=V_G  \cup V_J$, $E':=E_G \cup E_J$, and 
\[\lab'(u) \ := \ \left\{\begin{array}{ll}
\lab_G(u) & \mbox{if } u \in V_G\\
\lab_J(u) & \mbox{if } u \in V_J\\
\end{array} \right.\]
for every  $u \in V'$ is in $\CW_k$.
\item
Let $a,b \in [k]$ be two distinct integers and $G=(V_G,E_G,\lab_G) \in
\CW_k$ be a labeled graph, then
\begin{enumerate}
\item $\rho_{a \rightarrow b}(G):=(V_G,E_G,\lab')$ defined by 
\[\lab'(u) \ := \ \left\{\begin{array}{ll}
\lab_G(u) & \mbox{if } \lab_G(u) \not= a\\
        b & \mbox{if } \lab_G(u) = a\\
\end{array} \right.\]
for every $u \in V_G$ is in $\CW_k$ and
\item
$\eta_{a,b}(G)\ :=\ (V_G,E',\lab_G)$ defined by
$$E':=E_G \cup \{ \{u,v\} \mid u,v \in V_G,~u\not=v,~\lab(u)=a,~\lab(v)=b \}$$
is in $\CW_k$.
\end{enumerate}
\end{enumerate}
The {\em clique-width} of a labeled graph $G$, $\cws(G)$ for
short, is the least integer $k$ such
that $G \in \CW_k$.

An expression $X$ built with the operations
$\bullet_a,\oplus,\rho_{a \rightarrow b},\eta_{a,b}$ for integers $a,b \in
[k]$ is called a {\em clique-width $k$-expression}.
If integer $k$ is known from the context or irrelevant for the discussion, then we sometimes
use the simplified notion {\em expression} for the notion $k$-expression.
The graph defined by expression $X$ is denoted by $\graph(X)$.
Every unlabeled graph $G=(V,E)$ is considered as the labeled 
graph $(V,E,\lab)$ where $\lab : V \to [1]$.
\end{definition}

\begin{figure}[ht]
\begin{center}
\epsfig{figure=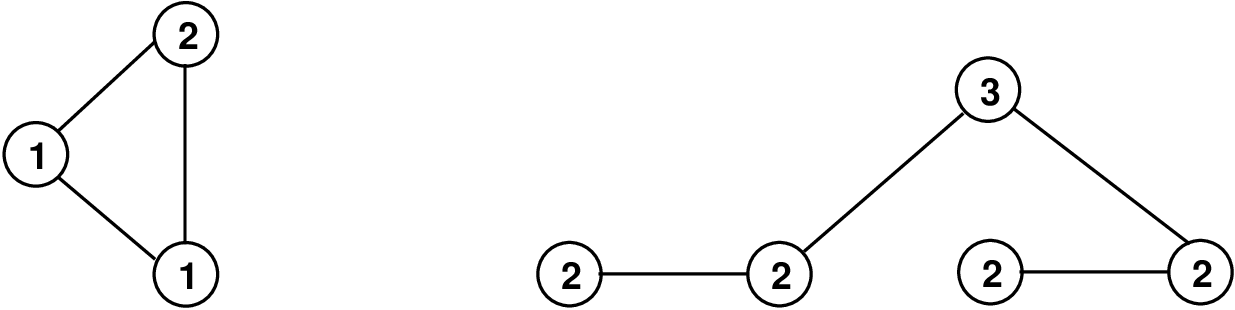,width=10.2cm}
\end{center}
\caption{Two labeled graphs $G_1$ and $G_2$ defined by expressions $X_1$ and $X_3$ and
by expressions $X_2$ and $X_4$, respectively.}
\label{Fgr}
\end{figure}

\begin{example}[Clique-width expressions]
The following two clique-width expressions $X_1$ and $X_2$ define the labeled graphs
$G_1$ and $G_2$ in Fig.~\ref{Fgr}.
\[X_1=\eta_{1,2}( (\rho_{2 \to 1}(\eta_{1,2}(\bullet_1 \oplus \bullet_2)))
\oplus \bullet_2)\]
\[X_2= \rho_{1 \to 2}(\eta_{2,3}
(((\eta_{1,2}(\bullet_1 \oplus \bullet_2))  \oplus (\eta_{1,2}(\bullet_1 \oplus \bullet_2)))
\oplus \bullet_3))\]
\end{example}

Since the clique-width edge insertion operations can be arranged in several ways, it is sometimes
useful to restrict to special clique-width expressions. 
\begin{itemize}
\item
A clique-width expression $X$ is  {\em irredundant}, if
for every subexpression  $\eta_{a,b}(X')$ of $X$,  in the graph 
$\val(X')$ no vertex labeled by $a$ is adjacent to a vertex labeled by $b$.
In \cite{CO00} it is shown that every graph which can be defined by
a clique-width $k$-expression can also be defined by an irredundant
clique-width $k$-expression.
\item
A clique-width expression $X$ is  {\em separated}, if
for every subexpression  $X'\oplus X''$ of $X$,  
the set of labels of the graph defined by $X'$ is disjoint from the set of labels of 
the graph defined by $X''$.
Every clique-width $k$-expression can be transformed into
an equivalent separated clique-width $2k$-expression, see \cite{CO00}.
\end{itemize}

\paragraph{NLC-width}
The notion of NLC-width\footnote{The abbreviation NLC results from the 
{\em node label controlled} embedding mechanism originally defined for graph
grammars.} of labeled graphs is defined by Wanke in \cite{Wan94} as follows.
\begin{definition}[NLC-width, \cite{CO00}]
Let $k$ be some positive integer. The class $\NLC_k$ of labeled graphs is
recursively defined as follows.
\begin{enumerate}
\item
The single vertex graph $\bullet_a$ for some $a \in [k]$ is in $\NLC_k$.
\item
Let $G=(V_G,E_G,\lab_G) \in \NLC_k$ and $J=(V_J,E_J,\lab_J) \in \NLC_k$ be
two vertex-disjoint  labeled graphs and $S \subseteq [k]^2$ be a relation, then
$$G \times_S J := (V',E',\lab')$$ defined by $V':=V_G \cup V_J$,
$$E':=E_G \cup E_J \cup \{\{u,v\} \mid u \in V_G,~v \in
V_J,~(\lab_G(u),\lab_J(v)) \in S \},$$ and  
\[\lab'(u) \ := \
\left\{\begin{array}{ll} \lab_G(u) & \mbox{if } u \in V_G\\
\lab_J(u) & \mbox{if } u \in V_J\\
\end{array} \right.\]
for every  $u \in V'$ is in $\NLC_k$.
\item
Let $G=(V_G,E_G,\lab_G) \in \NLC_k$ and $R:[k] \to [k]$ be a function, then
$$\circ_R(G) := (V_G,E_G,\lab')$$ defined by  
$$\lab'(u) := R(\lab_G(u))$$ for every  $u \in V_G$ is in $\NLC_k$.
\end{enumerate}
The {\em NLC-width} of a labeled graph $G$, $\nlcws(G)$ for
short, is the least integer $k$ such
that $G \in \NLC_k$.

An expression $X$ built with the operations
$\bullet_a,\times_S,\circ_R$ for $a \in [k]$, $S \subseteq [k]^2$, and $R:[k] \to [k]$ 
is called an {\em NLC-width $k$-expression}. 
If integer $k$ is known from the context or irrelevant for the discussion, then we sometimes
use the simplified notion {\em expression} for the notion $k$-expression.
The graph defined by expression $X$ is denoted by $\graph(X)$.
Every unlabeled graph $G=(V,E)$ is considered as the labeled 
graph $(V,E,\lab)$ where $\lab : V \to [1]$.
\end{definition}

\begin{example}[NLC-width expressions]
The following two NLC-width expressions $X_3$ and $X_4$ define the labeled graphs
$G_1$ and $G_2$ in Fig.~\ref{Fgr}.
\[X_3= (\bullet_1 \times_{\{(1,1)\}} \bullet_1)\times_{\{(1,2)\}} \bullet_2\]
\[X_4=\circ_{\{(1,2),(2,2),(3,3)\}}(((\bullet_1 \times_{\{(1,2)\}}\bullet_2)\times_{\emptyset}
(\bullet_1 \times_{\{(1,2)\}}\bullet_2))\times_{\{(2,3)\}}\bullet_3)\]
\end{example}

In contrast to clique-width expressions, NLC-width expressions are
always irredundant.

\paragraph{Expression Trees}
Every NLC-width $k$-expression $X$ has by its recursive definition a tree structure that is
called the {\em NLC-width $k$-expression-tree} for $X$.
This tree $T$ is an ordered rooted tree whose
leaves correspond to the vertices of graph $\val(X)$ and the inner 
nodes\footnote{To distinguish between the vertices of (non-tree) graphs and trees, 
we simply call the vertices of trees {\em nodes}.} correspond to
the operations of $X$, see \cite{GW00}.
In the same way we  define the clique-width $k$-expression-tree for every clique-width 
$k$-expression, see \cite{EGW03}.
If integer $k$ is known from the context or irrelevant for the discussion, then we sometimes
use the simplified notion {\em expression-tree} for the notion $k$-expression-tree.
For some node $u$ of expression-tree $T$, let $T(u)$ be the subtree of $T$ rooted at $u$.
Note that tree $T(u)$ is always an expression-tree. The expression $X(u)$ defined by
$T(u)$ can simply be determined by traversing the tree $T(u)$ starting from the root, where the
left children are visited first.  $X(u)$ defines a (possibly) relabeled induced 
subgraph $G(u)$ of $G$.
For an inner node $v$ of some expression-tree $T$ and a leaf $u$ of $T(v)$ we 
define by $\lab(u,G(v))$ the label  of that vertex of graph $G(v)$ that 
corresponds to $u$.
A node $u$ of  $T$ is called a {\em predecessor} of a node $u'$ of $T$
if $u'$ is on a path from $u$ to a leaf. A node $u$ of $T$ is called the {\em 
least common predecessor} of two nodes $u_1$ and $u_2$ if
$u$ is a predecessor of both nodes $u_1$, $u_2$, and no child of $u$ is a 
predecessor of $u_1$, $u_2$.

\paragraph{Graph Parameters and Relations}
There is a very close relation between the clique-width and the
NLC-width of a graph. We denote
two expressions $X_1$ and $X_2$ as {\em equivalent}, if the unlabeled
versions of $\val(X_1)$ and $\val(X_2)$ are isomorphic.

\begin{theorem}[\cite{Joh98}]\label{cw-nlcw}
Every clique-width $k$-expression can be transformed into an  equivalent
NLC-width $k$-expression and every NLC-width $k$-expression can be transformed 
into an  equivalent 
clique-width $2k$-expression. Thus, for every graph $G$ it holds
\begin{equation}
\nlcws(G)\leq \cws(G)\leq 2\cdot \nlcws(G).\label{eq-nlc-cw}
\end{equation}
\end{theorem}

In this paper we also refer to the notion of 
tree-width\footnote{The concept of tree-width already appeared 
in a work of Halin \cite{Hal76}.} of a graph $G$,  $\tws(G)$ for
short, which was defined in the 1980s 
by Robertson and Seymour in  \cite{RS86} by the existence of a 
tree-decomposition and to the notion of rank-width of  a graph $G$,  $\rws(G)$ for
short, which was introduced by Oum and Seymour in \cite{OS06}.

\begin{theorem}[Proposition 6.3 of \cite{OS06}]\label{cw-rw}
Every clique-width $k$-expression can be transformed into an  equivalent
rank-width $k$-expression and every rank-width $k$-expression can be transformed 
into an  equivalent 
clique-width $2^{k+1}-1$-expression. Thus, for every graph $G$ it holds
\begin{equation}
\rws(G)\leq \cws(G)\leq 2^{\rws(G)+1}-1.\label{eq-nlc-cw-rw1}
\end{equation}
\end{theorem}

The proof idea of Proposition 6.3 in \cite{OS06} immediately leads the following
bounds for NLC-width. The upper bound is lower, since NLC-width allows creating edges 
between equally labeled vertices.

\begin{theorem}\label{nlcw-rw}
Every NLC-width $k$-expression can be transformed into an equivalent
rank-width $k$-expression and every rank-width $k$-expression can be transformed 
into an  equivalent NLC-width $2^{k}$-expression. Thus, for every graph $G$ it holds
\begin{equation}
\rws(G)\leq \nlcws(G)\leq 2^{\rws(G)}.\label{eq-nlc-cw-rw2}
\end{equation}
\end{theorem}

\section{Binary Graph Operations and Graph Transformations}\label{sec-bin}

Let $G_1=(V_{G_1},E_{G_1})$ and $G_2=(V_{G_2},E_{G_2})$ be two non-empty
graphs and let $f$ be some binary graph operation which creates a new graph 
$f(G_1,G_2)$ from $G_1$ and $G_2$.  In this section we consider the NLC-width 
and clique-width of graph $f(G_1,G_2)$ with respect to the NLC-width or 
clique-width of $G_1$ and $G_2$.

\subsection{Disjoint Union}\label{section-disj-union}

The {\em disjoint union} of two vertex-disjoint graphs 
$G_1$ and $G_2$, denoted by  $G_1 \oplus G_2$, is the graph with vertex 
set $V_{G_1} \cup V_{G_2}$ and edge set $E_{G_1} \cup E_{G_2}$. 
Since NLC-width and clique-width operations both contain the disjoint union
it is easy to see that  
\[\nlcws(G_1 \oplus G_2)=\max(\nlcws(G_1),\nlcws(G_2))\] 
and
\[\cws(G_1 \oplus G_2)=\max(\cws(G_1),\cws(G_2)).\]

These bounds imply that the NLC-width and clique-width of a graph can be 
computed by the maximum NLC-width or clique-width of its connected 
components.

\subsection{Join}\label{section-join}

The {\em join}   of two vertex-disjoint graphs $G_1$ and $G_2$, denoted by $G_1 \otimes G_2$, 
is the graph with vertex set $V_{G_1} \cup V_{G_2}$ and edge set  
$$E_{G_1} \cup E_{G_2} \cup \{\{v_1,v_2\}~|~v_1\in V_{G_1},v_2\in V_{G_2}\}.$$
It is obviously that
\[\nlcws(G_1 \otimes G_2)=\max(\nlcws(G_1),\nlcws(G_2))\] 
and 
\[\cws(G_1 \otimes G_2)=\max(\cws(G_1),\cws(G_2),2).\]

Since NLC-width does not change when building the edge complement graph
(cf. Section \ref{EC}) we conclude that the NLC-width of a graph also can 
be computed by the maximum NLC-width  of its co-connected components, i.e.
the connected components of the edge complement graph.

\subsection{Union}\label{sec-sum}

The {\em union} of two graphs $G_1$ and $G_2$ with $V_{G_1}=V_{G_2}$, 
denoted by  $G_1 \cup G_2$, is the graph defined by the edge $E_{G_1} \cup E_{G_2}$.
Thus two vertices are adjacent in $G_1 \cup G_2$ if and only if they are adjacent in 
$G_1$ or they are adjacent in $G_2$.

Let $G_1$ be the disjoint union of $m$ paths $P_n$, each represented by a 
row in the adjacency matrix for $G_1$, and $G_2$ be the disjoint union of $n$
paths $P_m$, each represented by a column in the adjacency
matrix for $G_2$. Then the union $G_1 \cup G_2$ is an $n \times m$ grid.
Since paths have clique-width at most $3$ and 
an $n\times m$-grid has clique-width at least $\min(n,m)+1$ \cite{GR00},
it is not possible to bound the clique-width of $G_1 \cup G_2$ 
in the clique-width of $G_1$ and $G_2$, even if $G_1$ and $G_2$ are of 
bounded tree-width.

\subsection{Substitution}\label{Substi-oper}

Let $G_1$ and $G_2$ be two vertex-disjoint graphs and let $v\in V_{G_1}$ a vertex. 
The {\em substitution} of $v$  by $G_2$ in $G_1$, denoted by  $G_1[v/G_2]$, is the
graph with vertex set $V_{G_1}\cup V_{G_2} - \{v\}$ 
and edge set 
$$E_{G_1}\cup E_{G_2} - \{\{v,w\} ~|~ w\in N_{G_1}(v)\} \cup \{\{u,w\}~|~u\in V_{G_2},  
w\in N_{G_1}(v)\}.$$

Next we consider the NLC-width and clique-width of graph $G_1[v/G_2]$.

\begin{theorem}\label{th-substitution}
Let $G_1$ and $G_2$ be two vertex-disjoint graphs and $v\in V_{G_1}$ a vertex, 
then it holds
\[\nlcws(G_1[v/G_2])=\max(\nlcws(G_1),\nlcws(G_2))\] 
and 
\[\cws(G_1[v/G_2])=\max(\cws(G_1),\cws(G_2)).\] 
\end{theorem}

\begin{proof}
Let $G_1$ be a graph of NLC-width $k_1$, $v\in V_{G_1}$ a vertex, 
and $G_2$ be a graph of NLC-width $k_2$.
Let $T_1$ be an NLC-width $k_1$-expression-tree for $G_1$ and
$T_2$ be an NLC-width $k_2$-expression-tree for $G_2$.
Next we construct from $T_1$ and $T_2$ an expression-tree $T$ for $G_1[v/G_2]$.
We start with a copy $T$ of $T_1$. Let $x$ be the leaf of $T$ that
corresponds to vertex $v$.
We relabel $x$ from $\bullet_\ell$ into $\circ_R$, $R(a)=\ell$ for $a\in [k_2]$.
Then we insert a copy of $T_2$ in $T$ and make the root of
the copy of $T_2$ adjacent to leaf $x$ of $T$.
The resulting tree is an expression-tree for $G_1[v/G_2]$ using $\max(k_1,k_2)$
labels.

The clique-width result can be shown in the same way, see 
Lemma 3.4 in \cite{CO00}. 
\end{proof}

Vertex set $V_{G_2}$ is also called a {\em module} of the
graph $G_1[v/G_2]$, since all vertices of $V_{G_2}$ 
have the same neighbors in the graph $G_1[v/G_2]$. 
The substitution operation and quotient 
operation (cf. Section \ref{sec-sg})  are used in \cite{Joh98} and \cite{CMR00}
to show that the NLC-width and clique-width of a graph can be obtained by 
the maximum NLC-width or clique-width of  its prime
subgraphs appearing as quotient graphs in a modular decomposition.

\subsection{Product}

A graph product of two vertex-disjoint graphs $G_1$ and $G_2$ is a new graph 
whose vertex set is $V_{G_1} \times V_{G_2}$ and for two vertices $(u_1,u_2)$ 
and $(v_1,v_2)$ the adjacency in the product is defined by the adjacency, equality, or 
non-adjacency of $u_1$ and $v_1$ in $G_1$ and of $u_2$ and $v_2$ in $G_2$. 
Several results on graph products can be found in \cite{Har69,IK00,JT94}.
We consider the well known possibilities to define graph products shown
in Table \ref{tab-gp}.

\begin{table}[ht]
$$
\begin{array}{|l|l|}
\hline
\text{Graph product}    & \text{Edge set}=\{\{(u_1,u_2),(v_1,v_2)\} ~|~ \\
\hline
\text{Cartesian}      &(u_1=v_1 \wedge \{u_2,v_2\}\in E_{G_2}) \vee  (u_2=v_2 \wedge \{u_1,v_1\}\in E_{G_1})\}   \\
\text{Categorical}    & \{u_1,v_1\} \in E_{G_1}  \wedge \{u_2,v_2\} \in E_{G_2}\}  \\
\text{Normal}         &  (u_1=v_1 \wedge \{u_2,v_2\} \in E_{G_2}) \vee \\
&(\{u_1,v_1\} \in E_{G_1} \wedge  u_2=v_2) \vee\\
&(\{u_1,v_1\} \in E_{G_1}  \wedge \{u_2,v_2\} \in E_{G_2})\}\\
\text{Co-Normal}      &  \{u_1,v_1\} \in E_{G_1}  \vee  \{u_2,v_2\} \in E_{G_2}\} \\
\text{Lexicographic}  & (\{u_1,v_1\} \in E_{G_1}) \vee (u_1=v_1 \wedge \{u_2,v_2\} \in E_{G_2})\} \\
\hline
\end{array}
$$
\caption{Graph products}\label{tab-gp}
\end{table}

The cartesian, categorical, normal, and 
co-normal graph product applied to two paths $P_n$ and $P_m$ yields
a graph whose clique-width cannot be bounded independently
from $n$ and $m$. Thus it is not possible to bound the clique-width 
of the cartesian, categorical, normal, or 
co-normal graph product in the clique-width of the involved graphs.

The lexicographic graph product, which is also known as {\em graph composition},
of two graphs $G_1$ and $G_2$ is denoted by $G_1[G_2]$. 
Let $G^{0}=G_1$ and $V_{G_1}=\{v_1,\ldots,v_n\}$. 
Then  $$G^{i}=G^{i-1}[v_i/G_2], ~~i=1,\ldots,n$$ 
is a sequence of $n$ substitutions, such that $G^n$
defines graph $G_1[G_2]$. Thus we can apply Theorem \ref{th-substitution} 
to obtain the following results.

\begin{corollary}Let $G_1$ and $G_2$ be two vertex-disjoint graphs, then it holds
\label{Tlex}
\[\nlcws(G_1[G_2]) = \max(\nlcws(G_1),\nlcws(G_2))\] 
and
\[\cws(G_1[G_2]) = \max(\cws(G_1),\cws(G_2)).\]
\end{corollary}

\subsection{1-Sum}

Let $G_1$ and $G_2$ be two vertex-disjoint graphs and let $v\in V_{G_1}$  
and $w\in V_{G_2}$. The {\em 1-sum} of $G_1$ and $G_2$, denoted by
$G_1\oplus_{v,w}G_2$, consists of 
the disjoint union of $G_1$ and $G_2$ in which the two vertices  $v$ and $w$ are 
identified. That is, the graph $G_1\oplus_{v,w}G_2$ has vertex set 
$V_{G_1}\cup V_{G_2} - \{v,w\}\cup \{z\}$ and edge set 
$$
\begin{array}{lcl}
E_{G_1}\cup E_{G_2} &-&  \{\{v,v_1\}\in E_{G_1}~|~v_1 \in V_{G_1}\} \\
&-& \{\{w,w_1\}\in E_{G_2}~|~w_1 \in V_{G_2}\} \\
&\cup&\{\{z,z_1\}~|~z_1\in N_{G_1}(v)\cup N_{G_2}(w)\}.
\end{array}
$$

Next we consider the NLC-width and clique-width of graph $G_1\oplus_{v,w}G_2$.

\begin{theorem}\label{T1s} 
Let $G_1$ and $G_2$ be two vertex-disjoint graphs, $v\in V_{G_1}$ be a vertex, 
and $w\in V_{G_2}$ be a vertex. For $m_1=\max(\nlcws(G_1),\nlcws(G_2))$ 
it holds
\[m_1\leq \nlcws(G_1\oplus_{v,w}G_2)\leq m_1+1\] 
and  for $m_2=\max(\cws(G_1),\cws(G_2))$ it holds 
\[m_2\leq \cws(G_1\oplus_{v,w}G_2)\leq m_2+1.\] 
\end{theorem}

\begin{proof}
Let $G_1$ be a graph of NLC-width $k_1$, $v\in V_{G_1}$ a vertex, $G_2$ be 
a graph of NLC-width $k_2$, and $w\in V_{G_2}$ a vertex.
Let $T_1$ be an NLC-width $k_1$-expression-tree for $G_1$ and
$T_2$ be an NLC-width $k_2$-expression-tree for $G_2$.
We now construct an expression-tree $T$ for the graph $G_1\oplus_{v,w}G_2$
from $T_1$ and $T_2$, which uses $m_1+1$ labels.

We start with a copy $T$ of $T_2$. Let $x$ be the leaf of $T$ that corresponds to 
the vertex $w$. We relabel $x$ to $\bullet_{m_1+1}$ in order to substitute 
the vertex $w$ by the vertex $z$.
Now we consider all union nodes $x_1$ on the path from $x$ to the root of $T$ in $T$.
If $x$ is a left (right) child of $x_1$ and union node $x_1$ is labeled by $\times_S$ and
$(\lab(x,G(x_1)),\ell)\in S$ ($(\ell,\lab(x,G(x_1)))\in S$) for some $\ell\in[k_2]$
then we relabel  $x_1$ by $\times_{S'}$, where 
$S'=S\cup\{(m_1+1,\ell)~|~ (\lab(x,G(x_1)),\ell) \in S,~\ell\in[k_2]\}$    
($S'=S\cup\{(\ell,m_1+1)~|~ (\ell,\lab(x,G(x_1))) \in S,~\ell\in[k_2]\})$.
This is done in order to make in $G_1\oplus_{v,w}G_2$ all vertices adjacent to $z$ 
which are adjacent to $w$ in $G_2$.

We insert a new root $r$ labeled by $\circ_R$ and an edge from $r$ to the old root of $T$
into $T$. The relabeling $R$ maps every label from $[m_1]$ to $m_1+1$ and label
$m_1+1$ to $\ell$, if the leaf $y$ in $T_1$ which corresponds to vertex $v$ is labeled
by $\bullet_{\ell}$. Formally we have $R:[m_1+1]\to [m_1+1]$ and
$R(a)=m_1+1$ if $1\leq a\leq m_1$ and $R(a)=\ell$ if $a=m_1+1$. 

Further we insert a copy of $T_1$ in $T$ and replace the leaf $y$ by the root $r$. 
The labeling $\ell$ for the vertex $z$ ensures that   
all vertices which are adjacent to $v$ in $G_1$ become adjacent to $z$ in 
$G_1\oplus_{v,w}G_2$. 
The new root of $T$ is the root of $T_1$. Now $T$ defines the graph $G_1\oplus_{v,w}G_2$.

Since $G_1$ and $G_2$ are induced subgraphs of $G_1 \oplus_{v,w} G_2$, 
the NLC-width of $G_1 \oplus_{v,w} G_2$ is at least the maximum of 
the values $\nlcw(G_1)$ and $\nlcw(G_2)$.

In the same way we can show the clique-width result. The only difference is that we
have to ensure a non-used label to realize the relabeling operation. We can assume that
$m_2>1$ (otherwise $\cws(G_1\oplus_{v,w}G_2)=1$) and thus there is some
label $\ell'\in[m_2]$, $\ell'\neq \ell$, if the leaf $y$ in $T_1$ which corresponds to vertex 
$v$ is labeled by $\bullet_{\ell}$. Then the relabeling of tree $T$ where $w$ is labeled
by $m_2+1$ can be done as follows. First we map all labels from $[m_2]$ to $\ell'$,
then we map label $m_2+1$ to $\ell$, and finally we map label $\ell'$ to $m_2+1$. The obtained
tree can be glued to tree $T_1$ as described above.
\end{proof}

The shown NLC-width bounds are tight for $m_1=1$ and $m_1=2$, which can be
verified by the 1-sums $P_2\oplus_{v,w}P_3$ and $P_5\oplus_{v,w}P_6$, where
$v$ and $w$ are vertices of degree $1$ within the involved paths.

If $v$ and $w$ in the definition of the 1-sum are not isolated vertices in
$G_1$ and $G_2$ the new vertex $z$ 
is also called an {\em articulation vertex} of the graph $G_1\oplus_{v,w}G_2$, 
since $G_1\oplus_{v,w}G_2$ without $z$
has more connected components than $G_1\oplus_{v,w}G_2$. The maximal connected 
subgraphs of some graph $G$ without any articulation vertex are called {\em blocks} 
of $G$. The bounds of Theorem \ref{T1s} imply that the NLC-width and clique-width 
of a graph  can be bounded by the maximum NLC-width or clique-width of its blocks and 
its number of blocks. By a deeper analysis in  \cite{BL02,LR04b} it has been shown that
the clique-width 
of a graph  can be bounded by the maximum clique-width of its blocks plus $2$, which
implies that every graph of clique-width $k$
contains a block whose clique-width is at least $k-2$.

\subsection{Corona}\label{sec-corona}

The corona of graphs was introduced by Frucht and Harary in \cite{FH70},  
when constructing a graph whose automorphism group is the wreath product of 
the two component automorphism groups.
The {\em corona}  of two vertex-disjoint graphs $G_1$ and $G_2$, denoted by $G_1 \wedge G_2$,
consists of the disjoint union of one copy of $G_1$ and $|V_{G_1}|$ copies of $G_2$ and
each vertex of the copy of $G_1$ is connected to all vertices of one copy of $G_2$, i.e.
$|V_{G_1}|\cdot|V_{G_2}|$ edges are inserted in the disjoint union of the $|V_{G_1}|+1$ 
graphs.

The corona of $G_1$ and $G_2$ can also be obtained by applying 1-sum operations as follows.
Let $V_{G_1}=\{v_1,\ldots,v_n\}$ be the vertex set of $G_1$. For $i=1\ldots,n$ we
take a copy of $G_2$ and insert a dominating vertex $w_i$ (cf. Section \ref{vertexadd}) to that copy 
and obtain a graph  $G_{2,i}$. Then by the sequence of 1-sums
$$(\ldots((G_1\oplus_{v_1,w_1}G_{2,1})\oplus_{v_2,w_2}G_{2,2})\ldots)\oplus_{v_n,w_n}G_{2,n}$$
we obtain the corona $G_1 \wedge G_2$.
By this observation, we can bound the NLC-width and the clique-width of $G_1 \wedge G_2$ in 
the NLC-width or the clique-width of its combined graphs by applying the idea of
the proof of Theorem \ref{T1s} on every leaf of an expression-tree for $G_1$.

\begin{theorem}\label{Tcor}Let $G_1$ and $G_2$ be two vertex-disjoint graphs. 
Further let $m_1=\max(\nlcws(G_1),$ $\nlcws(G_2))$, then it holds 
\[m_1\leq \nlcws(G_1 \wedge G_2) \leq m_1 +1\] 
and for $m_2=\max(\cws(G_1),\cws(G_2))$ it holds 
\[m_2\leq \cws(G_1 \wedge G_2) \leq m_2 +1.\]
\end{theorem}

\section{Unary Graph Operations and Graph Transformations}\label{sec-un}

Let $G=(V_G,E_G)$ be a non-empty graph and  $f$ be some unary graph transformation 
which creates a new graph $f(G)$ from $G$. In this section we consider the NLC-width 
and clique-width of the graph $f(G)$ with respect to the NLC-width or 
clique-width of graph $G$.

\subsection{Vertex Deletion and Vertex Addition\label{vertexadd}}

\paragraph{Vertex Deletion}
Let $G$ be a graph and $v\in V_G$. By $G-v$ we denote the graph which we 
obtain from $G$ by removing vertex $v$ and all edges incident to $v$.
That is,
$$G-v=(V_G-\{v\},E_G - \{\{v,v'\}~|~v'\in N(v)\}).$$

Next we consider the NLC-width and clique-width of graph $G-v$.

\begin{theorem}
\label{Tvertexrem}
Let $G$ be a graph and $v\in V_G$, then it holds
\[\nicefrac{1}{2}\cdot \nlcws(G) \le\nlcws(G-v)\le \nlcws(G)\]
and
\[\nicefrac{1}{2}\cdot\cws(G) \le\cws(G-v)\le  \cws(G). \]
\end{theorem}

\begin{proof} 
An NLC-width $k$-expression-tree and also a clique-width $k$-expression-tree 
for the graph $G-v$ can be obtained 
by a $k$-expression-tree $T$ for the graph $G$ by removing the leaf $x$ corresponding
to vertex $v$ and some obvious cleaning of the tree because operations at 
predecessors of $x$ lost one input.

Since we can obtain $G$ by inserting $v$ into $G-v$ 
Theorem \ref{Tvertex} leads the lower bounds.
\end{proof}

\paragraph{Vertex Addition}
Let $G$ be a graph, $N\subseteq V_G$, and $v\not\in V_G$. By $G+_N v$ 
we denote the graph which we 
obtain from $G$ by inserting vertex $v$ with 
neighborhood $N(v)=N$. That is,
$$G+_N v=(V_G\cup\{v\},E_G\cup \{\{v,v'\}~|~v'\in N\}).$$
In the special case where $N(v)=\{v'\}$ for some $v'\in V_G$ 
we call $v$  a {\em pendant vertex}
and where $N(v)=V_G$ we call $v$ a {\em dominating vertex}.

Next we consider the NLC-width and clique-width of  graph $G+_N v$.

\begin{theorem}
\label{Tvertex}
Let $G$ be a graph, $N\subseteq V_G$, and $v\not\in V_G$, then it holds
\[\nlcws(G) \le\nlcws(G+_N v)\le 2\cdot \nlcws(G)\]
and
\[\cws(G) \le\cws(G+_N v)\le 2\cdot \cws(G). \]
\end{theorem}

\begin{proof} 
Let $G$ be a graph of NLC-width $k$, $N\subseteq V_G$, $v\not\in V_G$ be a vertex, and 
$T$ be an NLC-width $k$-expression-tree that defines the graph $G$.
We now define an NLC-width $2k$-expression-tree that defines the graph $G+_N v$.
We start with a copy $T'$ of $T$.

First we separate the neighborhood of $v$ from the non-neighborhood by 
introducing $k$ further labels $k+1,\ldots,2k$.
Every leaf of $T'$ that corresponds to a vertex from $G$ which is not from $N$ 
will be relabeled from label $\bullet_a$, $a\in[k]$, into $\bullet_{a+k}$. 

Then we consider all nodes $x$ on the paths from these
relabeled leaves to the root of the so defined tree. If node $x$ is a union node  
labeled by some $\times_S$, $S\subseteq [k]^2$, then we relabel $x$ by $\times_{S'}$ where 
$S'=\{(a,b),(a,b+k),(a+k,b),(a+k,b+k) ~|~(a,b)\in S\}$.
If node $x$ is a relabeling node labeled by
some $\circ_R$, $R: [k]\to [k]$, then we relabel $x$ by 
$\circ_{R'}$, where $R':[2k]\to[2k]$ and $R'(a)=R(a)$, 
if $i\leq k$ and  $R'(a)=R(a)+k$, if $k+1\leq a\leq 2k$. The resulting tree
is denoted by $T''$.

In a last step we insert two additional nodes $t_v$ and $t_r$ 
labeled by  $\bullet_1$ and $\times_{\{(1,a)~|~a\in [k]\}}$,
respectively  and two additional arcs from  $t_r$ to  $t_v$ 
and from $t_r$ to the root of $T''$ in $T''$, such that $t_v$ is
the left child of $t_r$.

The resulting tree is denoted by $T'''$. The tree $T'''$ is an NLC-width 
$2k$-expression-tree and $T'''$ defines the graph $G+_N v$.

Since we can obtain $G$ by removing $v$ from $G+_N v$, 
Theorem \ref{Tvertexrem} leads the lower bounds.

\medskip
To prove the corresponding clique-width bound,  we can construct 
a clique-width $2k$-expression-tree $T''$ which defines the same graph 
as the tree $T''$ defined above using the same ideas as for NLC-width. 
Then we have to find a label for vertex $v$, 
which is not used in the graph defined by $G(T'')$, since
clique-width does not allow edge insertions between equal labeled vertices. This
can be done by relabeling all vertices $G(T'')$ labeled by $k+1,\ldots,2k$ by e.g. 
$k+1$ and then we can take, for $k\ge 2$, one of the free labels e.g. label $2k$ to 
label the inserted vertex $v$. In the case $k=1$, $G$ consists of isolated vertices
and $G+_N v$ is the disjoint union of one $K_{1,p}$, for some $p$, and isolated vertices. 
Thus also in this case $G+_N v$ has clique-width $2k=2$.
\end{proof}

The shown NLC-width bounds are tight for graphs of width $1$ and $2$.
If we insert a vertex in a path of length $2$ to get a path of length $3$, we
insert a vertex in a graph of $\nlcw$ $1$ and obtain a graph of $\nlcw$ $2$.
If we insert vertex $v$ in the graph $H-v$ of Fig.~\ref{Ff}, we
insert a vertex in a graph of $\nlcw$ $2$ and obtain a graph of $\nlcw$ $4$.

Since the addition of a dominating vertex will be used in several of our 
constructions (cf. Section \ref{sec-corona} and Section \ref{sec-sc}) we state the following result.

\begin{corollary}\label{cor-dv}
Let $G$ be a graph and $v\not\in V_G$, then it holds
\[\nlcws(G+_{V_G} v) =\nlcws(G)\]
and
\[\cws(G+_{V_G} v)=\max(\cws(G),2).\]
\end{corollary}

Further it is possible to bound the NLC-width and clique-width
of $G+_N v$ in the NLC-width and clique-width of $G$ and the vertex degree $d=|N|$
of $v$. The main idea is to label each vertex of $G$ which should
be adjacent to vertex $v$ by a new label  from $\{k+1,\ldots,k+d\}$.
Then, the new vertex can easily be inserted in a last step. If we
use clique-width operations we first have to relabel at least one
of the used labels from $\{1,\ldots,k\}$ to get a free label in order to 
insert the new vertex.

\begin{corollary}
\label{Tvertexd}
Let $G$ be a graph, $N\subseteq V_G$, $d=|N|$, and $v\not\in V_G$, 
then it holds
\[\nlcws(G) \le\nlcws(G+_N v)\le \nlcws(G)+d \]
and
\[\cws(G) \le\cws(G+_N v)\le \cws(G)+d. \]
\end{corollary}

The addition of a vertex of high degree $d'=|V_G|-d$ can be done more 
efficiently by adding a vertex of degree $d$ in the edge complement
and building the edge complement of the result. By the NLC-width bound of
Section \ref{EC} we get the following result.

\begin{corollary}
\label{Tvertexd2}
Let $G$ be a graph, $N\subseteq V_G$, $d=|V_G|-|N|$, and $v\not\in V_G$, 
then it holds
\[\nlcws(G) \le\nlcws(G+_N v)\le \nlcws(G)+d.\]
\end{corollary}

For clique-width the latter approach does not lead a better bound 
than that of Theorem \ref{Tvertex}.

\subsection{Edge Addition and Edge Deletion}\label{edgeadd}

Let $G$ be a graph and $v,w\in V_G$ two vertices. For $\{v,w\}\not\in E_G$ we define by
$G+\{v,w\}$ the graph we obtain from $G$ by adding edge $\{v,w\}$. That is,
$$G+\{v,w\}=(V_G,E_G\cup\{\{v,w\}\}).$$ 
For $\{v,w\}\in E_G$ we define by
$G-\{v,w\}$ the graph we obtain from $G$ by deleting edge $\{v,w\}$. That is,
$$G-\{v,w\}=(V_G,E_G-\{\{v,w\}\}).$$

Our next theorem shows that we can insert or delete an edge in a graph
using at most $2$ more labels.

\begin{theorem}
\label{Tedge}
Let $G$ be a graph and $v,w\in V_G$ be two different vertices, then  it holds
\[\nlcws(G) - 2\le\nlcws(G\pm\{v,w\})\le \nlcws(G) + 2\]
and
\[\cws(G) - 2\le\cws(G\pm\{v,w\})\le \cws(G) + 2.\]
\end{theorem}

\begin{proof}
In order to show the upper bound on NLC-width,
let $G$ be a graph of NLC-width $k$ and let $v$ and $w$ be two non-adjacent
vertices of $G$. Further, let $T$ be an NLC-width $k$-expression-tree that defines $G$.
We now define an NLC-width $(k+2)$-expression-tree that defines $G+\{v,w\}$.
We start with a copy $T'$ of $T$. Let $x$ and $y$ be the leaves of $T'$ that
correspond to vertices $v$ and $w$, respectively, of the graph $G$. First, we relabel leaf
$x$ and $y$ in $T'$ by $\bullet_{k+1}$ and $\bullet_{k+2}$, respectively.

Next we consider all union nodes $x_1$ on the path from $x$ to the root of $T'$ in $T'$.
If $x$ is a left (right) child of $x_1$ and union node $x_1$ is labeled by $\times_S$ and
$(\lab(x,G(x_1)),\ell)\in S$ ($(\ell,\lab(x,G(x_1)))\in S$) for some $\ell\in[k]$
then we relabel  $x_1$ by $\times_{S'}$, where 
$S'=S\cup\{(k+1,\ell)~|~ (\lab(x,G(x_1)),\ell) \in S,~\ell\in[k]\}$    
($S'=S\cup\{(\ell,k+1)~|~ (\ell,\lab(x,G(x_1))) \in S,~\ell\in[k]\})$.
This is done in order to make all vertices which are adjacent to 
$v$ in $G$ also adjacent to $v$ in $G+\{v,w\}$.
In the same way we modify all union nodes on the path from $y$ to the root of $T'$
in order to preserve the adjacencies of $w$ in $G+\{v,w\}$.

Last we relabel the least  common predecessor $z$ of $x$ and $y$ in $T'$ to
create the edge between $v$ and $w$. Since $z$ is always a union node in $T'$, $z$ is labeled
by $\times_S$ for some $S\subseteq [k]^2$. If $x$ is the left (right) child and $y$ is the right
(left) child of $z$ in $T'$ then we relabel $z$ by $\times_{S\cup\{(k+1,k+2)\}}$
($\times_{S\cup\{(k+2,k+1)\}}$).

The resulting tree is denoted by $T''$. The tree $T''$ is an NLC-width
$(k+2)$-expression-tree and $T''$ defines the graph $G+\{v,w\}$.

The proof for edge deletion runs similar, we just have to leave out the above described 
relabeling of the least common predecessor  $z$ of $x$ and $y$ in $T'$ to create 
the edge between $v$ and $w$.

For the lower bounds assume that $H$ is obtained from $G$ by inserting (deleting) an 
edge $e$ and it holds $\nlcw(H)< \nlcw(G)-2$. Then by deleting (inserting) edge $e$ 
from (in) $H$ we obtain some graph $G'$ which is isomorphic to $G$. By our shown upper bound we
conclude that $\nlcw(G')< \nlcw(G)$, which leads to a contradiction.

The results for clique-width can be shown by similar arguments.
\end{proof}

If we add or delete an edge in a graph of NLC-width 1, i.e. a co-graph,
then we always obtain a graph of NLC-width at most 2, since we
can label both end vertices of the new edge (both end vertices of the deleted edge) 
by the same label 2.

\begin{figure}[ht]
\begin{center}
\epsfig{figure=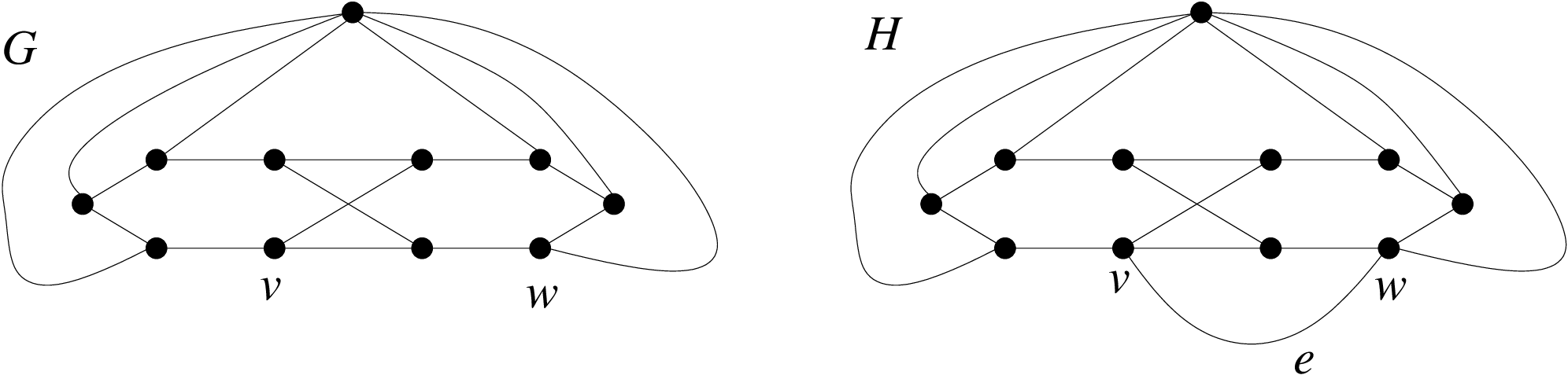,width=14cm}
\end{center}
\caption []{\label{Ff}The graph $G$ has $\nlcw$ 2. 
The graph $H$ can be obtained from $G$
by adding edge $e$ and $H$ has $\nlcw$ 4.}
\end{figure}

The graphs in Fig.~\ref{Ff} can be used to show that the NLC-width bounds of Theorem
\ref{Tedge} cannot be improved for $k=2$. For the edge addition we observe that 
the graph $G$ has NLC-width $2$ and
the graph $H$, which we obtain after inserting edge $e=\{v,w\}$ in $G$,
has NLC-width $4$ which was found by a computer program.\footnote{We implemented
an algorithm which takes as input a graph $G$ and an integer $k$ and which decides
whether $\nlcws(G)\leq k$.}
For the edge deletion we notice that the complement graph $\overline{G}$ of $G$ has 
NLC-width $2$ and contains edge $\{v,w\}$.
If we remove edge $\{x,y\}$ from $\overline{G}$ we obtain the graph $\overline{H}$
which has NLC-width $4$.

Our last theorem gives an answer of Question 6.3 of \cite{CO00}, which asks
how different the clique-width of two graphs can be if they differ by exactly
one edge. It remains to verify whether the given clique-width bounds of 
Theorem \ref{Tedge} are tight.

\begin{problem}
Is there a graph $G$ and $v,w\in V_G$, such that $|\cws(G)-\cws(G+\{v,w\})|=2$?
Is there a graph $H$ and $\{v,w\}\in E_H$, such that $|\cws(H)-\cws(H-\{v,w\})|=2$?
Or can the results on the clique-width of Theorem \ref{Tedge} be improved?
\end{problem}

\subsection{Edge Subdivision}

Let $G$ be a graph, $u\not\in V_G$, and  $\{v,w\}\in E_G$. The {\em subdivision}  
of $\{v,w\}$ in $G$, $Subdiv(G,v,w)$ for
short, 
has vertex set $V_G\cup\{u\}$ and edge set $E_G-\{\{v,w\}\} \cup \{\{v,u\},\{w,u\}\}$. 
The subdivision operation is also known as {\em elementary refinement}.

Next we analyze the effect of an edge subdivision 
on the  NLC-width and clique-width of a given graph.

\begin{theorem}
\label{Tsub}
Let $G$ be a graph and $\{v,w\}\in E_G$ an edge, then it holds
\[ \nlcws(G) - 2 \le\nlcws(Subdiv(G,v,w))\le \nlcws(G) + 2\]
and
\[\cws(G) - 2\le\cws(Subdiv(G,v,w))\le \cws(G) + 2.\]
\end{theorem}

\begin{proof}
First we want to show the upper bound for NLC-width.
Let $G$ be a graph of NLC-width $k$ and let $\{v,w\}$ be an edge of $G$.
Let $T$ be an NLC-width $k$-expression-tree that defines $G$. 
We now define an NLC-width $(k+2)$-expression-tree that defines $Subdiv(G,v,w)$.

Let $T'$ be defined for $T$ as in the proof of Theorem \ref{Tedge}
for edge removing. In $T'$ we insert a new root $r$ labeled by 
$\times_{\{(k+1,k+1),(k+2,k+1)\}}$
and a new node $z$ (defining the vertex $u$ which subdivides edge $\{v,w\}$) 
labeled by $\bullet_{k+1}$ and
two edges, one from $r$ to $z$ and one from $r$ to the root of $T'$ 
such that $z$ is the right child of $r$.

The resulting tree is denoted by $T''$. Then $T''$ is an NLC-width
$(k+2)$-expression-tree and it is easy to show that $T''$ defines the graph $Subdiv(G,v,w)$.

For the lower bounds assume that the graph $Subdiv(G,v,w)$ is obtained from $G$ by subdividing 
an edge $\{v,w\}$ and $\nlcw(Subdiv(G,v,w))< \nlcw(G)-2$. Then we obtain by removing 
the inserted vertex $u$ and inserting $\{v,w\}$
in $Subdiv(G,v,w)$ a graph $G'$ isomorphic to graph $G$ with $\nlcw(G')< \nlcw(G)$, by 
our upper bound in Theorem \ref{Tedge}, and thus a contradiction.

Since clique-width operations do not allow edge insertions between equal labeled
vertices, we have to do one additional relabeling $\rho_{k+1\to k+2}$ in order to 
label vertices $v$ and $w$ in the proof of Theorem \ref{Tedge} by $k+2$ before 
inserting the new vertex in $T$.
\end{proof}

The upper bound for $\nlcw(Subdiv(G,v,w))$ 
of Theorem \ref{Tsub} cannot be improved, since first subdividing an edge and deleting the 
new vertex corresponds to edge deletion, which needs two additional labels in general,
see Fig.~\ref{Ff}.

In the appendix of \cite{CO00} it is shown that in a graph $G$  of clique-width 
at least $4$ every path of length at least 5, consisting of vertices which all 
have degree $2$ in $G$ and one end vertex of degree $1$ in $G$, can be extended by 
subdivisions without increasing the clique-width of $G$.

There are several examples where a subdivision increases 
NLC-width and clique-width, e.g. a $P_3$, and several 
examples where a subdivision does not change
NLC-width and clique-width, e.g. a $P_4$. It remains
open, whether a subdivision can decrease the NLC-width
and clique-width of graphs.

\begin{problem}
Is there a graph $G$ and an edge $\{v,w\}\in E_G$, such that  
$\nlcws(Subdiv(G,v,w))<\nlcws(G)$ or $\cws(Subdiv(G,v,w))<\cws(G)$?
\end{problem}

At least after subdividing all edges of a graph the resulting graph 
is bipartite. If we subdivide every edge of a graph $G$ we obtain the  
so-called {\em incidence graph} $I(G)$ of the graph $G$. 
Incidence graphs have unbounded clique-width in general, but incidence graphs
of graphs of bounded tree-width have bounded clique-width, since subdivisions
do not change the tree-width. 
The following very close bound has been shown in \cite{Bou14}.
\begin{equation}
\cws(I(G)) \leq   \tws(G)+3 \label{ic}
\end{equation}
Since there exist graphs of tree-width $k$ and clique-width at least
$2^{\lfloor \frac{k}{2} \rfloor-1}$ by \cite{CR05}, the transformation
from $I(G)$ to $G$ can increase clique-width exponentially.
Further applications of bound (\ref{ic}) can be found in \cite{Cou15}.

\subsection{Vertex Identification and Edge Contraction}

For some graph $G$ and two different vertices $v,w\in V_G$ the {\em identification} of
$v$ and $w$ in $G$, $Ident(G,v,w)$  for short, has vertex set $V_G-\{v,w\}\cup\{u\}$ 
and edge set 
\[\begin{array}{lcl}
E_G&- & \{\{v',v''\}~|~ v'\in V_G, v''\in\{v,w\}\}\\
&\cup& \{\{v',u\}~|~ v'\not\in\{v,w\} \text{ and }\{v',v\}\in E_G \text{ or }  \{v',w\}\in E_G\}.
\end{array}
\]
Next we analyze the identification of two vertices 
in a graph with respect to the NLC-width and clique-width of the involved graphs.

\begin{theorem}\label{Tid}
Let $G$ be a graph and $v,w\in V_G$, then it holds 
\[\nicefrac{1}{4}\cdot \nlcws(G) \le\nlcws(Ident(G,v,w))\le 2\cdot \nlcws(G)\]
and
\[\nicefrac{1}{4}\cdot \cws(G) \le\cws(Ident(G,v,w))\le 2\cdot \cws(G).\]
\end{theorem}

\begin{proof}
For the upper bound we can delete $v,w$ and insert $u$ 
with neighborhood $N(v)\cup N(w)$ (cf. Theorem \ref{Tvertex}).
The lower bound holds since we can obtain $G$ from $Ident(G,v,w)$ by removing $u$ and inserting 
the vertices $v$ and $w$, each with a factor of 2 (cf. Theorem \ref{Tvertex}).
\end{proof}

If the two vertices $v$ and $w$ of an identification are adjacent, i.e. $\{v,w\}\in E_G$, 
we call the corresponding operation {\em edge contraction}, which is a well known minor 
operation. Courcelle has shown in \cite{Cou14} that there is a graph of
clique-width 3, which yields a graph of clique-width greater than $3$ by
the contraction of a single edge. This disproves Conjecture 4.4 in 
\cite{LPRW12} on the closure of graphs of bounded clique-width 
under edge contractions.

In the appendix of \cite{CO00} it is shown that in a graph $G$  of 
clique-width at least $4$ every path of length at least $2$, consisting 
of vertices which all have degree $2$ 
in $G$ and one end vertex of degree $1$ in $G$, can be decreased by 
edge contractions without increasing the clique-width of $G$.

\subsection{Subgraph}\label{sec-sg}

\paragraph{Subgraph}
For an arbitrary subgraph $H$ of a graph $G$, and thus also for 
an arbitrary minor, the clique-width of $H$ and NLC-width of $H$ 
cannot be bounded in the clique-width or NLC-width of $G$. This can easily 
be shown by the example of complete graphs, which all have NLC-width $1$ and 
clique-width $2$, while their subgraphs may have arbitrary large  NLC-width 
and clique-width. By taking the number of removed edges into account,
the bounds of Section \ref{edgeadd} can be used to estimate the NLC-width
and clique-width of subgraphs.

\paragraph{Induced Subgraph}
Since every induced subgraph $H$ of a graph $G$ can be
realized by vertex deletions, by Section \ref{vertexadd} it holds
\[\nlcws(H)\leq \nlcws(G)\] 
and 
\[\cws(H)\leq \cws(G).\]
Although taking induced subgraphs does not increase the NLC-width and clique-width
of a graph, characterizations for the classes $\NLC_k$, $k\ge 2$, and 
$\CW_k$, $k\ge 3$, by sets of forbidden induced subgraphs
are unknown until now.

\paragraph{Quotient}
If we remove all but one vertices of a module $V'\subseteq V_G$ from graph $G$,
we denote the obtained graph as a {\em quotient graph} of $G$.  
Since every quotient graph of $G$   is an induced subgraph of 
$G$, the quotient operation does not increase NLC-width or clique-width.

\subsection{Power of a Graph}

The {\em $d$-th power} $G^d$ of a graph $G$ is a graph with the same set of 
vertices as $G$ and an edge between two vertices if and only if there is a 
path of length at most $d$ between them.
Suchan and Todinca have shown  in \cite{ST07}  the following bound.
\[\nlcws(G^d) \le 2\cdot(d+1)^{\nlcws(G)} \]

\subsection{Line Graph}

The {\em line graph} $L(G)$ of a graph $G$ has a vertex for every edge of $G$
and an edge between two vertices if the corresponding edges of $G$ are 
adjacent \cite{Whi32}. For some line graph $L(G)$, the graph $G$ is called the
{\em root graph} of $L(G)$.
Even for complete graphs $K_n$, the line graph operation generates
graphs whose NLC-width  cannot be bounded in the NLC-width of their
root graphs \cite{GW07b}. But it is possible to bound the NLC-width and clique-width
of line graphs in the tree-width of their root graphs, and even vice versa by the
following bounds, which have been shown in  \cite{GW07b}.
\[
\nicefrac{1}{4}\cdot (\tws(G)+1) \leq \nlcws(L(G)) \leq  \tws(G)+2
\]
\[
\nicefrac{1}{4}\cdot (\tws(G)+1)  \leq  \cws(L(G)) \leq 2\cdot\tws(G)+2
\]

\subsection{Edge Complement}\label{EC}

The {\em edge complement graph} $\overline{G}$ of a graph $G$ has the same 
vertex set as $G$ and two vertices in $\overline{G}$ are adjacent if and 
only if they are not adjacent in $G$, i.e. 
$$\overline{G}=(V_G,\{\{u,v\}~|~u,v\in V_G, u\neq v, \{u,v\}\not\in E_G\}).$$
The following bounds and proof ideas are known from \cite{Wan94} and \cite{CO00}.

\begin{theorem} Let $G$ be a graph, then
\label{Th-edge-c}
\[\nlcws(\overline{G})= \nlcws(G)\]
and
\[\nicefrac{1}{2}\cdot\cws(G)\le\cws(\overline{G})\leq 2\cdot\cws(G).\]
\end{theorem}

\begin{proof}
Let $T$ be an NLC-width $k$-expression-tree that defines the graph $G$. 
We now define a new NLC-width $k$-expression-tree that defines the graph $\overline{G}$.  
Let $T'$ be a copy of $T$. Every node labeled by $\times_S$ in $T'$  
is relabeled by $\times_{S'}$, where $S'=\{(a,b)~|~(a,b)\not\in S,~a,b\in[k]\}$. 
Finally tree $T'$ is an NLC-width $k$-expression-tree and defines graph $\overline{G}$.
Since the complement of the  complement graph is the original graph, the
claimed equality holds true.

Let $G$ be a graph of clique-width $k$.
In order to show the upper bound on the clique-width of graph $\overline{G}$
we assume that we have given a separated $2k$-expression for $G$ (cf. Section 
\ref{intro} and \cite{CO00}), which allows to exchange the existing edges
by the non-existing edges. As above, since the complement of the complement 
graph is the original graph, the lower bound follows.
\end{proof}

\subsection{Bipartite Complement}

Let $G$ be a bipartite graph with vertex partition $V_G=V_1 \cup V_2$, 
such that there are no edges between two vertices of $V_1$ and no edges 
between two vertices of $V_2$. The {\em bipartite complement}
$\overline{G}^{\text{bip}}$ of $G$ has  the same 
vertex set as $G$ and its edge set is obtained by complementing the
edges between $V_1$ and $V_2$, i.e.
$$\overline{G}^{\text{bip}}=(V_G,
\{\{u,v\} ~|~ \{u,v\}\not\in E_G, u\in V_1, v\in V_2\}).$$
The following clique-width bound is known from \cite{LR04}.

\begin{theorem} Let $G$ be a bipartite graph, then
\label{Th-edge-bc}
\[\nicefrac{1}{2}\cdot\nlcws(G)\le\nlcws(\overline{G}^{\text {bip}})\leq 2\cdot\nlcws(G)\]
and
\[\nicefrac{1}{4}\cdot\cws(G) \le\cws(\overline{G}^{\text {bip}})\leq 4\cdot\cws(G).\]
\end{theorem}

\begin{proof}
Let $G=(V,E)$ be a bipartite graph of clique-width $k$ and $T$ be
a clique-width $k$-expression-tree for $G$. By Theorem \ref{Th-edge-c}
there is a clique-width $2k$-expression-tree $T'$ for graph $\overline{G}$.
If we denote the bipartition $G$ by $V_1\cup V_2$, then we have to
choose from $\overline{G}$ only those edges where one vertex is from $V_1$ and
one vertex is from $V_2$. Therefore in \cite{LR04} for every label
$i\in[2k]$ two labels $i_1$ and $i_2$ for the vertices in $V_1$ 
and $V_2$ are introduced. We modify the nodes $x$ in $T'$ as follows.
\begin{enumerate}
\item If $x$ is a leaf labelled by $\bullet_i$ corresponding to a vertex
from $V_1$ then we relabel $x$ by $\bullet_{i_1}$ and if $x$ is 
a leaf labelled by $\bullet_i$ corresponding to a vertex
from $V_2$ then we relabel $x$ by $\bullet_{i_2}$.

\item If $x$ represents a relabeling operation $\rho_{i\to j}$ and $y$ 
is the direct predecessor of $x$, 
then we relabel  $x$ by $\rho_{i_1\to j_1}$,
insert a further node $x'$ labelled by 
$\rho_{i_2\to j_2}$  into $T'$, and two new arcs from $y$ to $x'$ and
from $x'$ to $x$.

\item If $x$ represents an edge insertion operation $\eta_{i,j}$ 
and $y$ is the direct predecessor of $x$, 
then we  relabel  $x$ by
$\eta_{i_1,j_2}$  and insert a further node $x'$ labelled by $\eta_{i_2,j_1}$ 
into $T'$, and two new arcs from $y$ to $x'$ and
from $x'$ to $x$.

\end{enumerate}
This leads a clique-width $4k$-expression-tree for graph $\overline{G}^{\text {bip}}$.
Since the bipartite complement of the bipartite complement graph is the original graph, 
the lower bound follows.

The NLC-width bounds can be obtained even easier, since by Theorem \ref{Th-edge-c}
there is an NLC-width $k$-expression-tree $T'$ for graph $\overline{G}$. Thus we can
obtain an NLC-width $2k$-expression-tree for graph $\overline{G}^{\text {bip}}$.
\end{proof}

\subsection{Local Complementation}

For some graph $G$ and  a vertex $v\in V_G$ the {\em local complementation} 
$LC(G,v)$ is defined by Bouchet in \cite{Bou94} as follows. The graph $LC(G,v)$ 
is obtained from the graph $G$ by replacing the subgraph of $G$ defined by $N(v)$ 
by its edge complement, i.e. $LC(G,v)$ has vertex set $V_G$ and edge set
$$
\begin{array}{lcl}
E_G &   - & \{\{u,w\}~|~u,w\in N_G(v), \{u,w\}\in E_G\} \\
    &\cup &\{\{u,w\}~|~u,w\in N_G(v), u\neq w, \{u,w\}\not\in E_G\}.
\end{array}
$$

In Corollary 2.7 in \cite{Oum05a} it is shown that
the rank-width of a graph does not change 
by applying local complementations, which leads to a characterization 
of graphs of rank-width at most $k$ by finitely many
forbidden vertex-minors (i.e. taking induced subgraphs and 
local complementations). 

Next we consider the NLC-width and clique-width of graph $LC(G,v)$.

\begin{theorem} Let $G$ be a graph and $v\in V_G$, then
\label{Tlc}
\[\nicefrac{1}{2}\cdot\nlcws(G) \le\nlcws(LC(G,v))\le 2\cdot \nlcws(G)\]
and
\[\nicefrac{1}{3}\cdot\cws(G) \le\cws(LC(G,v))\le 3\cdot \cws(G).\]
\end{theorem}

\begin{proof}
Let $T$ be an NLC-width $k$-expression-tree that defines the graph $G$. We now define
a new NLC-width $2k$-expression-tree that defines the graph $LC(G,v)$.  We start with
a copy $T'$ of $T$. The main idea is to separate the labels of the vertices in
$N(v)$ from the labels of the vertices in $V-N(v)$. Let $n'=|N(v)|$ and
$x_1,\ldots,x_{n'}$ be the leaves of $T'$ that corresponds to vertices in  
$N(v)$ of $G$.

For every leaf $x_i$, $i=1,\ldots,n'$, we modify the nodes $x$ on
the paths from $x_i$ to the root of $T'$ in $T'$ as follows.
\begin{enumerate}
\item If $x$ is a leaf $x_i$, $i=1,\ldots,n'$, 
labeled by $\bullet_\ell$ in $T'$, then we relabel $x$ by $\bullet_{\ell+k}$.

\item If $x$ is a relabeling node labeled by $\circ_R$, 
then we relabel  $x$ by $\circ_{R'}$, such that $R'(a)=R(a)$, if $1\leq a \leq k$ 
and $R'(a)=R(a-k)+k$, if $k+1 \leq a \leq 2k$.

\item If $x$ is a union node labeled by $\times_S$, then we relabel $x$ 
by $\times_{S'}$, such that
$S'=S\cup S_1 \cup S_2$, where $S_1=\{(a+k,b+k)~|~(a,b)\not\in S\}$ and
$S_2=\{(a,b+k),(a+k,b)~|~(a,b)\in S\}$.
Set $S_1$ creates an edge between two vertices in $N(v)$, if and only if 
these vertices are not adjacent in $G$ and set $S_2$ creates an edge between 
one vertex of $V_G-N(v)$ and one vertex of $N(v)$, if and only if these 
vertices are adjacent in $G$.
\end{enumerate}

These three steps create the complement graph of the subgraph 
induced by $N(v)$. The resulting tree is denoted by $T''$. The tree $T''$ is an 
NLC-width $2k$-expression-tree and defines graph $LC(G,v)$.

The lower bound follows since by $L(L(G,v),v)$ we obtain $G$.

For the clique-width bounds we need $k$ additionally labels to distinguish 
the vertices in $N(v)$ from those in $V-N(v)$ and $k$ further labels to 
create the complement graph of the subgraph induced by vertex set $N(v)$.
\end{proof}

\begin{figure}[ht]
\begin{center}
\epsfig{figure=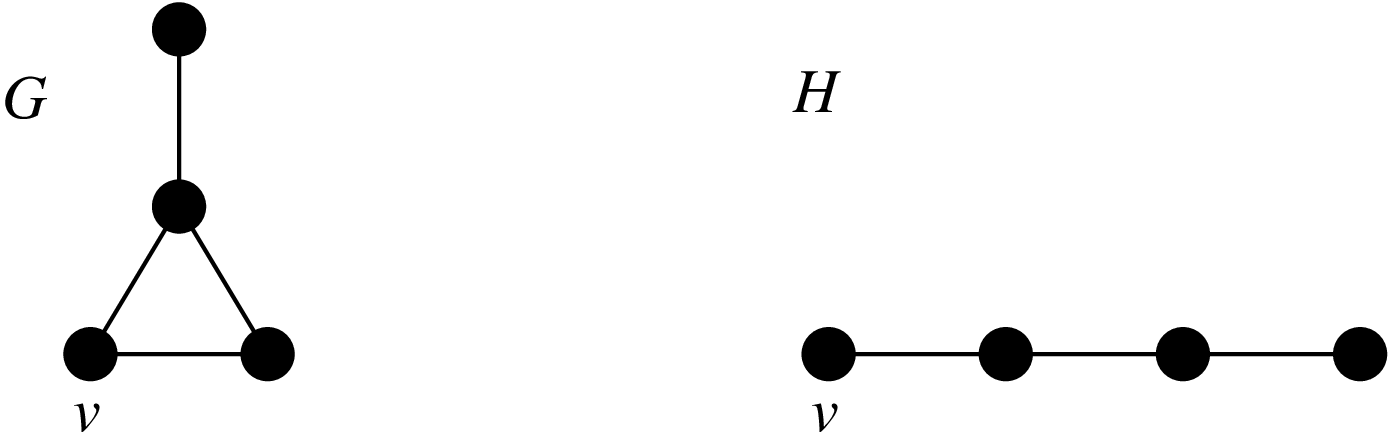,width=7.2cm}
\end{center}
\caption{\label{Fg}The graph $G$ on the left side
and has $\nlcw$ $1$ ($\cw$ $2$). The graph $H$ on the right side  has 
$\nlcw$ $2$ ($\cw$ $3$).}
\end{figure}

The graph $G$ given in Fig.~\ref{Fg} 
(which is called {\em paw} or {\em 3-pan} in \cite{BLS99}) 
shows that the local complementation
can increase or decrease the NLC-width and clique-width of a graph by 1. 
If we apply a local complementation on one of the vertices of degree $2$ in $G$, we obtain 
a path on four vertices.

The proof of Theorem \ref{Tlc} implies the following bounds 
for the NLC-width and clique-width of the graph $LC(G,v)$
using the vertex degree of $v$ in the  graph $G$.

\begin{corollary}Let $G$ be a graph and $v\in V_G$, then
\label{Tlcx}
\[\nlcws(LC(G,v))\le  \nlcws(G) + \min(\nlcws(G),\deg_G(v))\]
and 
\[\cws(LC(G,v))\le \cws(G) + 2\cdot \min(\cws(G),\deg_G(v)).\]
\end{corollary}

Two graphs $G$ and $G'$ on the same vertex set are called 
{\em locally equivalent} if there is a sequence of vertices 
$(v_1,\ldots,v_\ell)$ such that $G^0=G$, $G^i=LC(G^{i-1},v_i)$ for 
$i=1,\ldots,\ell$ and $G^\ell=G'$.

\begin{theorem}\label{loc-eq}
Let $G$ be a graph and $G'$ a graph which
is locally equivalent to $G$, then it holds
\label{Tlcr}
\[\nlcws(G')\le 2^{\nlcws(G)}\]
and
\[\cws(G')\le 2^{\cws(G)+1}-1.\]
\end{theorem}

\begin{proof}
To show the clique-width bound let $G$ be a graph of clique-width $k$.
By Theorem \ref{cw-rw} we know that $G$ has rank-width at most $k$.
Since the rank-width of a graph does not change 
by applying local complementations (cf. Corollary 2.7 in \cite{Oum05a}), 
every graph $G'$ which is  obtained by a sequence of local complementations on $G$ also has 
rank-width at most $k$. Applying Theorem \ref{cw-rw}, we know that $G'$ has
clique-width at most $2^{k+1}-1$. 

The NLC-width bound follows in the same way by Theorem \ref{nlcw-rw}.
\end{proof}

\subsection{Seidel Switching}

The {\em switching} operation is defined by Seidel 
in connection with regular structures, such as systems of equiangular
lines, strongly regular graphs, or the so-called two-graphs, 
see \cite{Sei74a,Sei76,ST81}. Several examples of applications of Seidel switching can be 
found in algorithms, e.g. in a polynomial-time algorithm for 
the $P_3$-structure recognition problem \cite{Hay96} and for the construction
of bi-join decomposition of graphs \cite{MR05}.
Let $G$ be a graph and $v\in V_G$ be a vertex. The graph $S(G,v)$ has 
the same vertex set as $G$ and its edge set is the edge set of $G$ 
but changing the neighbors of $v$ to non neighbors and vice versa.
That is,  the graph  $S(G,v)$ has vertex set $V_G$ and edge set
\[\begin{array}{lcl}
E_G & -    & \{\{v,w\}~|~w\in V_G, \{v,w\}\in E_G\} \\
    & \cup & \{\{v,w\}~|~w\in V_G,v\neq w, \{v,w\}\not\in E_G\}.
\end{array}
\]

Next we will show that one switching operation in a graph increases or 
decreases its NLC-width and clique-width by at most one.

\begin{theorem}\label{Tswitch}
Let $G=(V_G,E_G)$ be a graph and $v\in V_G$, then it holds
\[\nlcws(G)-1 \le\nlcws(S(G,v))\le  \nlcws(G)+1\]
and
\[\cws(G)-1 \le\cws(S(G,v))\le  \cws(G)+1.\]
\end{theorem}

\begin{proof}
Let $T$ be an NLC-width $k$-expression-tree that defines $G$ and $v\in V_G$. 
We now define a new NLC-width $(k+1)$-expression-tree that defines $S(G,v)$.  
We start with a copy $T'$ of $T$.
Let $x$ be the leaf of $T'$ that corresponds to vertex $v$ of $G$.
We relabel the leaf $x$ in $T'$ by $\bullet_{k+1}$.

Now we consider the union nodes $x_1$ on the path from $x$ to the root of $T'$ in $T'$.
If $x$ is a left (right) child of $x_1$ and union node $x_1$ is labeled by $\times_S$
then we relabel  $x_1$ by $\times_{S'}$, where
$S'=S\cup\{(k+1,\ell)~|~(\lab(x,G(x_1)),\ell)\not\in S, \ell\in[k]\}$
($S'=S\cup\{(\ell,k+1)~|~(\ell,\lab(x,G(x_1)))\not\in S, \ell\in[k]\}$).
This is necessary in order do make all vertices adjacent to $v$ which are not adjacent to
$v$ in $G$, and vice versa.

The resulting tree is denoted by $T''$. The tree $T''$ is 
an NLC-width $(k+1)$-expression-tree
and $T''$ defines the graph $S(G,v)$.

The lower bound follows since by $S(S(G,v),v)$ we obtain $G$.

In order to show the bound on the clique-width of graph $S(G,v)$
we assume that we have given an irredundant 
expression for $G$ (cf. Section \ref{intro}).
\end{proof}

The NLC-width bounds given in Theorem \ref{Tswitch} are best possible.
For the upper bound consider the graph $G$ of NLC-width $1$ in Fig.~\ref{Fg}.
A switching operation on the graph $G$ at one of the vertices of 
degree $2$ creates a graph $H$ which is isomorphic to a $P_4$, which has 
NLC-width $2$. Further by $S(H,v)$ we obtain the graph $G$, thus the lower 
bound is best possible too.

Two graphs $G$ and $G'$ on the same vertex set are called {\em switching equivalent} if 
there is a sequence of vertices $(v_1,\ldots,v_\ell)$ such
that $G^0=G$, $G^i=S(G^{i-1},v_i)$ for $i=1,\ldots,\ell$ and $G^\ell=G'$.
It is shown in \cite{CC80} that deciding if two graphs are switching equivalent 
is an isomorphism complete problem.

\begin{theorem} \label{the-sw}
Let $G$ be a graph and $G'$ a graph which
is switching equivalent to $G$ by sequence $(v_1,\ldots,v_\ell)$, then it holds
\[\nlcws(G')\le 2^{\nlcws(G)+\ell}\]
and
\[\cws(G')\le 2^{\cws(G)+\ell+1}-1.\]
\end{theorem}

\begin{proof}
Let $G=(V,E)$ be a graph of NLC-width $k$. In order to express a sequence $(v_1,\ldots,v_\ell)$
of $\ell$ switching operations by local complementations we insert $2\ell$ vertices $u_1,\ldots,u_{\ell}$
and $w_1,\ldots,w_{\ell}$ into $G$, such that $N(u_i)=V-\{v_i\}$ and $N(w_i)=V$.
The resulting graph $G'$ has NLC-width at most $k+\ell$ (cf. Section \ref{vertexadd}) and 
rank-width at most $k+\ell$ (cf. Theorem \ref{nlcw-rw}). 
Further the sequence of local complementations $(u_1,w_1,\ldots,u_\ell,w_\ell)$ on $G'$ 
creates a graph $G''$, which is isomorphic to the graph obtained by the sequence  
$(v_1,\ldots,v_\ell)$ of switching operations on graph $G$.
Since the rank-width of a graph does not change by applying local complementations 
(cf. Corollary 2.7 in \cite{Oum05a}), 
graph $G''$ also has rank-width at most $k+\ell$. By Theorem  \ref{nlcw-rw} we know that $G''$ 
has NLC-width at most $2^{k+\ell}$.

The clique-width result can be obtained using the same arguments but using 
Theorem  \ref{cw-rw} instead of Theorem  \ref{nlcw-rw}.
\end{proof}

\begin{problem}
Can we bound the NLC-width and clique-width of $G'$ in Theorem \ref{the-sw} independently
from the number of applied switching operations $\ell$? (For locally equivalent graphs
and Seidel complementation  equivalent graphs this is possible by Theorem \ref{loc-eq} and 
Theorem \ref{Tlsw1}.)
\end{problem}

\subsection{Seidel Complementation}\label{sec-sc}

The {\em Seidel complementation} operation is defined by Limouzy in \cite{Lim10}
in order to give a characterization for permutation graphs. 
Let $G$ be a graph and $v\in V_G$ be a vertex. The graph $SC(G,v)$ has 
the same vertex set as $G$ and its edge set is the edge set of $G$ 
but complementing the edges between the neighborhood and the non-neighborhood
of $v$. That is,  the graph  $SC(G,v)$ has vertex set $V_G$ and edge set
\[\begin{array}{lcl}
E_G \triangle \{\{x,y\} ~|~ \{v,x\}\in E_G, \{v,y\}\not\in E_G\},
\end{array}
\]
where $A\triangle B = (A-B)\cup (B-A)$ denotes the symmetric difference 
of two sets $A$ and $B$.

Next we consider the NLC-width and clique-width of graph $SC(G,v)$.

\begin{theorem}\label{Tswitch2}
Let $G=(V_G,E_G)$ be a graph and $v\in V_G$, then it holds
\[\nicefrac{1}{2}\cdot\nlcws(G)-1 \le\nlcws(SC(G,v))\le  2\cdot\nlcws(G)+1\]
and
\[\nicefrac{1}{2}\cdot\cws(G)-1\le\cws(SC(G,v))\le  2\cdot\cws(G)+1.\]
\end{theorem}

\begin{proof}
Let $T$ be an NLC-width $k$-expression-tree that defines the graph $G$. We now define
a new NLC-width $(2k+1)$-expression-tree that defines the graph $SC(G,v)$.  
We start with a copy $T'$ of $T$. The main idea is to separate the labels of the 
vertices in sets $\{v\}$, $N(v)$, and $V-(N(v)\cup\{v\})$ pairwise from each other. 

First we separate the label of vertex $v$.
Let $x_0$ be the leaf of $T'$ that corresponds to vertex $v$ of $G$.
We relabel the leaf $x_0$ in $T'$ by $\bullet_{2k+1}$.
Now we consider the union nodes $x$ on the path from $x_0$ to the root of $T'$ in $T'$.
If $x_0$ is a left (right) child of $x$ and union node $x$ is labeled by $\times_S$
then we relabel  $x$ by $\times_{S'}$, where
$S'=S\cup\{(2k+1,\ell)~|~(\lab(x,G(x)),\ell)\in S, \ell\in[k]\}$
($S'=S\cup\{(\ell,2k+1)~|~(\ell,\lab(x,G(x)))\in S, \ell\in[k]\}$).
By this process the adjacencies of $v$ do not change.

Next we separate the labels of the vertices in $V-(N(v)\cup\{v\})$ and complement 
the edges between the neighborhood and the non-neighborhood of $v$.
Let $n'=|V-(N(v)\cup\{v\})|$ and $x_1,\ldots,x_{n'}$ be the leaves of $T'$ that 
correspond to vertices in $V-(N(v)\cup\{v\})$ of $G$. 
For every leaf $x_i$, $i=1,\ldots,n'$, we modify the nodes $x$ on
the paths from $x_i$ to the root of $T'$ in $T'$ as follows.
\begin{enumerate}
\item If $x$ is a leaf $x_i$, $i=1,\ldots,n'$, 
labeled by $\bullet_\ell$ in $T'$, then we relabel $x$ by $\bullet_{\ell+k}$.

\item If $x$ is a relabeling node labeled by $\circ_R$, 
then we relabel  $x$ by $\circ_{R'}$, such that $R'(a)=R(a)$, if $1\leq a \leq k$ 
and $R'(a)=R(a-k)+k$, if $k+1 \leq a \leq 2k$.

\item If $x$ is a union node labeled by $\times_S$, then we relabel $x$ 
by $\times_{S'}$, such that
$S'=S\cup S_1 \cup S_2$, where $S_1=\{(a+k,b+k)~|~(a,b)\in S\}$ and
$S_2=\{(a,b+k),(a+k,b)~|~(a,b)\not\in S\}$.
Set $S$ creates an edge between two vertices in $N(v)$, 
set $S_1$ creates an edge between two vertices in $V-(N(v)\cup\{v\})$, and
set $S_2$ creates an edge between one vertex in $N(v)$ and one vertex in $V-(N(v)\cup\{v\})$, 
if and only if these vertices are not
adjacent in $G$.
\end{enumerate}
These three steps complement 
the edges between the neighborhood and the non-neighborhood of $v$.
The resulting tree is denoted by $T''$. The tree $T''$ is an 
NLC-width $(2k+1)$-expression-tree and defines graph $SC(G,v)$.

The lower bound follows since by $SC(SC(G,v),v)$ we obtain $G$.

In order to show the bound on the clique-width of graph $SC(G,v)$
we assume that we have given an irredundant expression for $G$ (cf. Section 
\ref{intro} and \cite{CO00}).
\end{proof}

Two graphs $G$ and $G'$ on the same vertex set are called {\em Seidel complementation  equivalent} if 
there is a sequence of vertices $(v_1,\ldots,v_\ell)$ such
that $G^0=G$, $G^i=SC(G^{i-1},v_i)$ for $i=1,\ldots,\ell$ and $G^\ell=G'$.

\begin{theorem}\label{Tlsw1} Let $G$ be a graph and $G'$ a graph which
is Seidel complementation equivalent to $G$, 
then it holds
\[\nlcws(G')\le 2^{\nlcws(G)}\]
and
\[\cws(G')\le 2^{\cws(G)+1}-1.\]
\end{theorem}

\begin{proof}
Let $G$ be a graph, $v\in V_G$ be a vertex, and $G'=SC(G,v)$.
Let $G_0$  be the graph obtained from $G$ by adding a dominating
vertex $v_0$ and $G'_0$  be the graph obtained from $G'$ by adding a dominating
vertex $v_0$. It is easy to check that $G'_0$ can be obtained from $G_0$ (up to isomorphism) 
by applying three local complementations\footnote{The
application of three local complementations at $v$, at $v_0$, and again at $v$ 
for some edge $\{v,v_0\}$ is also known as {\em pivoting} the edge $\{v,v_0\}$, 
see \cite{Oum05a}.} at $v$, at $v_0$, and again at $v$.
This implies that the rank-width of $G'_0$ and $G_0$ are equal (cf. Corollary 2.7 in \cite{Oum05a}).

Now, suppose that $G_1$ and $G_2$ are two Seidel complementation equivalent  graphs. Then
$G_{1,0}$ and $G_{2,0}$ (both obtained by adding a dominating vertex $v_0$)
are locally equivalent and therefore  $G_{1,0}$ and $G_{2,0}$ have the same
rank-width. Then the following estimations holds.
$$
\begin{array}{lclll}
\nlcws(G_{1})    & =    & \nlcws(G_{1,0})      && \text{by Corollary  \ref{cor-dv}}    \\
                 & \leq & 2^{\rws(G_{1,0})}    && \text{by Theorem \ref{nlcw-rw}}          \\
                 & =    & 2^{\rws(G_{2,0})}    && \text{by Corollary 2.7 in \cite{Oum05a} } \\
                 & \leq & 2^{\nlcws(G_{2,0})}  && \text{by Theorem \ref{nlcw-rw}}  \\
                 & =    & 2^{\nlcws(G_{2})}    && \text{by Corollary  \ref{cor-dv}} 
\end{array}
$$

Using Theorem \ref{cw-rw} instead of Theorem \ref{nlcw-rw} one can prove the
clique-width bound. Since graphs of clique-width $1$ are edgeless and for 
these graphs  a  Seidel complementation does not change the graph, we can
restrict to graphs of clique-width is at least 2, such that the 
addition of dominating vertices does not change the width by 
Corollary  \ref{cor-dv}.
\end{proof}

\section{Conclusions and Outlook}\label{sec-con}

We considered a number of binary graph transformations $f$
which create some new graph $f(G_1,G_2)$ from two graphs $G_1$ and $G_2$. 
In all cases in which it is possible to bound the NLC-width and
clique-width of the combined graph $f(G_1,G_2)$ in the NLC-width and
clique-width of graphs $G_1$ and $G_2$
we show how to compute the corresponding expression in linear time in
the size of the corresponding expressions for $G_1$ and $G_2$. 
Thus our results are constructive.
In Table \ref{Ta2} we  compare these results.

\begin{table}[ht]
\begin{center}
\begin{tabular}{|l|c|c|}
\hline
transformation $f$ & $\nlcws(f(G_1,G_2))$      &   $\cws(f(G_1,G_2))$  \\
\hline\hline
disjoint union     & $\max(k_1,k_2)$     & $\max(k_1,k_2)$       \\
join               & $\max(k_1,k_2)$     & $\max(k_1,k_2,2)$  \\
substitution       & $\max(k_1,k_2)$     & $\max(k_1,k_2)$   \\
composition        & $\max(k_1,k_2)$     & $\max(k_1,k_2)$   \\
1-sum              & $\max(k_1,k_2)+1$   & $\max(k_1,k_2)+1$   \\
corona             & $\max(k_1,k_2)+1$   & $\max(k_1,k_2)+1$   \\
\hline
\end{tabular}
\end{center}
\caption{Let $G_1$ and $G_2$ be two graphs of $\nlcw$  (or clique-width) $k_1$ and $k_2$, respectively,
and $f$ be a binary graph transformation  of the first column.
The second column of the table shows the upper bound of the $\nlcw$ of graph
$f(G_1,G_2)$. The third column gives the results for $\cw$.\label{Ta2}}
\end{table}


Furthermore we have shown how the NLC-width and clique-width of a given
graph change if we apply certain unary graph transformation $f$ on this graph.
In all cases in which it is possible to bound the NLC-width and clique-width 
of the resulting graph $f(G)$
we also show how to compute the corresponding expression in linear time in
the size of the corresponding expression for $G$. 
Although clique-width is the more famous concept, we obtain in all cases closer bounds
for $\nlcw(f(G))$ for local transformations $f$. 
In Table \ref{Ta1} we  compare our results concerning unary graph transformations.

\begin{table}[ht]
\begin{center}
\begin{tabular}{|l|c|c|}
\hline
transformation $f$ & $\nlcws(f(G))$    &  $\cws(f(G))$  \\
\hline\hline
vertex insertion     & $2k$  & $2k$     \\
edge insertion       & $k+2$ & $k+2$    \\
edge deletion        & $k+2$ & $k+2$  \\
edge subdivision     & $k+2$ & $k+2$      \\
edge contraction     & $2k$  & $2k$    \\
induced subgraph     & $k$   & $k$      \\
edge complement      & $k$   & $2k$    \\
bipartite complement & $2k$  & $4k$    \\
local complementation& $2k$  & $3k$    \\
switching            & $k+1$ & $k+1$   \\
Seidel complementation &  $2k+1$ &  $2k+1$   \\
\hline
\end{tabular}
\end{center}
\caption{Let $G$ be a graph of $\nlcw$ (or clique-width) $k$ 
and $f$ be a unary graph transformation of the first column.
The second column of the table shows the upper bound of the $\nlcw$ of graph
$f(G)$. The third column gives the results for $\cw$.\label{Ta1}}
\end{table}

Since the computation of NLC-width and clique-width is NP-hard \cite{GW07b,FRRS09},
it seems to be difficult to find an optimal $k$-expression for some given graph. 
Our results may help to find an expression for some graph of interest $f(G)$, if
we have an expression for graph $G$ and $f$ is one of the transformations
listed in Table \ref{Ta1}.
For example, we can construct an NLC-width 
$(k+\ell)$-expression for every graph  which is switching equivalent to some graph with 
known NLC-width $k$-expression, where $\ell$ is the number of necessary switching transformations.
As well, we can construct an $(k+2)$-expression for every graph which differs only by one
edge from a graph with known $k$-expression.

Our estimations can also be made for the clique-width of directed graphs, 
which was defined in  \cite{CO00} and for the 
NLC-width of directed graphs, which was defined in \cite{GWY16}.
In order to carry over the notations local complementation, switching, Seidel complementation, and
edge complement, we define 
for some directed graph $G=(V,E)$ its complement digraph by
$$\overline{G}=(V,\{(u,v)~|~(u,v)\not\in E, u,v\in V, u\neq v\}).$$
For the neighborhood of a vertex $v\in V$ 
the sets $N_G^+(v)=\{u\in V~|~ (v,u)\in E\}$,
$N_G^-(v)=\{u\in V~|~ (u,v)\in E\}$, and 
$N_G(v)=N_G^+(v)\cup N_G^-(v)$
can be chosen.
In this way all bounds of Tables \ref{Ta2} and \ref{Ta1} can be shown in 
the same way as done for the parameters on undirected graphs in this paper.

Furthermore linear clique-width and linear NLC-width, which
are defined in \cite{GW05a}, can be bounded when considering 
graph operations.
One  difference to the general versions of the parameters is 
that the linear NLC-width and
the linear clique-width do not allow the disjoint union or join of
two graphs on more than one vertex. Thus  for the  transformations listed in 
Table \ref{Ta2} the linear NLC-width and linear clique-width bounds 
for disjoint union rises to $\max(k_1,k_2)+1$ and
for join rises to $\max(k_1,k_2)+1$.
A further difference is
that the linear clique-width of $\overline{G}$ is at most 
linear clique-width of $G$ plus $1$ \cite{GW05a} while the
linear NLC-width does not change as known from the general
version.  
This implies that for the transformations listed in 
Table \ref{Ta1} the linear clique-width bounds 
for edge complement reduces to $k+1$,
for bipartite complement reduces to $2k+2$, and
for local complementation reduces to $2k+1$.
All other mentioned bounds of Tables \ref{Ta2} and \ref{Ta1}
can also be shown for the linear NLC-width and
the linear clique-width.

There are several open questions. 
In nearly all cases, it remains to show that our  bounds are best possible,
or to improve them. Especially the clique-width bounds on bipartite
complement and local complementation seem to be improvable.

Further it remains open if there are graph transformations (cf. Section \ref{sec-intro} 
for the definition), which do not increase
the clique-width or NLC-width of a given graph and make the given graph smaller, 
in order to define useful reduction rules or a characterization by
forbidden graphs for graphs of bounded clique-width or graphs of bounded NLC-width. 
Among our considered transformations only
the induced subgraph transformation does not increase the clique-width or NLC-width,
which implies that there exist characterizations by sets of forbidden induced graphs
for $\NLC_k$ and $\CW_k$ for every integer $k$. Unfortunately only for
$\NLC_1$ and $\CW_2$, i.e. the set of all co-graphs, these sets are known.
For the sets $\NLC_3$ there is no characterization
by a set of {\em finitely} many forbidden induced subgraphs, since every $n$-vertex 
cycle $C_n$ with $n\geq 11$ has NLC-width 4. The same holds for the set $\CW_3$, 
since every $n$-vertex cycle $C_n$ with $n\geq 7$ has clique-width 4.

It is also an open problem to find graph operations that increase or decrease the 
NLC-width or clique-width of
some graph by a fixed constant or a fixed factor, e.g. an operation such that for 
every graph $G$ there is a positive integer  $c$ such that $\nlcws(f(G))=c+\nlcws(G)$ or
$\nlcws(f(G))=c \cdot \nlcws(G)$.
This would imply a useful means in order to decrease NLC-width or clique-width
in a controlled way. For rank-width the transformation from $G=(V,E)$ into the
bipartite graph $B(G)=(V',E')$, where $V'=V\times \{1,2,3,4\}$ and
$$E'=\{\{(v,i),(v,i+1)\}~|~ v\in V, i\in[3]\} \cup \{\{(v,1),(w,4)\}~|~ \{v,w\}\in E\}$$
increases the width by a
factor of $c=2$, see Lemma 5.3 in \cite{Oum05b}.


\bibliographystyle{alpha}
\bibliography{/home/gurski/bib.bib}

\end{document}